\documentclass{mem}
\usepackage{natbib}\usepackage{txfonts}\usepackage{balance}
\usepackage{graphicx}
\usepackage[a4paper,breaklinks,dvipdfm]{hyperref}
\idline{75}{282}
\begin{document}
\def\teff{$T\rm_{eff }$}
\def\kms{$\mathrm {km s}^{-1}$}

\title{
A window on the efficiency of the s-process in AGB stars: chemical abundances of n-capture elements in the planetary nebula NGC\,3918
}

   \subtitle{}

\author{
S. \,Madonna\inst{1,2} 
\and J. \,Garc\'{\i}a-Rojas\inst{1,2} \and V. \,Luridiana\inst{1,2} \and N. \,C. \,Sterling\inst{3} \and C. \,Morisset\inst{4}
          }

\institute{
Instituto de Astrof\'{\i}sica de Canarias, E-38205 La Laguna, Tenerife, Spain
\and
Universidad de La Laguna, Dpto. Astrof\'{\i}sica, E-38206 La Laguna, Tenerife, Spain
\and
Department of Physics, University of West Georgia, 1601 Maple Street, Carrollton, GA 30118, USA
\and
Instituto de Astronom\'{\i}a, Universidad Nacional Aut\'onoma de M\'exico, Apdo. Postal 70264, M\'exico D. F. 04510, M\'exico
}

\authorrunning{Madonna et al. }

\titlerunning{s-process in NGC\,3918}

\abstract{
The chemical content of the planetary nebula NGC\,3918 is investigated through deep, high-resolution (R$\sim$40000) UVES at VLT spectrophotometric data. We identify and measure more than 750 emission lines, making ours one of the deepest spectra ever taken for a planetary nebula. Among these lines we detect very faint lines of several neutron-capture elements (Se, Kr, Rb, and Xe), which enable us to compute their chemical abundances with unprecedented accuracy, thus constraining the efficiency of the {\emph s}-process and convective dredge-up in the progenitor star of NGC\,3918. 

\keywords{Stars: AGB and post-AGB -- ISM: abundances -- ISM: planetary nebulae -- individual: NGC\,3918 }
}
\maketitle{}

\section{Introduction}

About half of the heavy elements (Z $>$ 30) in the Universe are formed during the AGB phase through slow neutron-capture ({\emph n}-capture) nucleosynthesis (the ``\emph{s}-process''). In the intershell between the H- and He-burning shells, Fe-peak nuclei experience neutron captures interlaced with $\beta$-decays, which transform them into heavier elements. In the same layer, carbon is formed and brought to the surface by the third dredge-up along with n-capture elements, to be successively expelled to the interstellar medium by stellar winds and planetary nebula (PN) ejection. 
The efficiency of nucleosynthesis and the third dredge-up in AGB stars can be effectively constrained by measuring the s-element abundances; but {\emph n}-capture elements are extremely underabundant and detecting their lines requires very deep spectra.

Although first identified in 1994 \citep{pequignotbaluteau94}, {\emph n}-capture element emission lines were not well-studied in ionized nebulae until recently. \citet{sharpeeetal07} carried out high-resolution observations of 4 PNe on 4- and 6-m class telescopes, identifying lines of Br, Kr, Xe, Rb, Ba, Te, and I ions. But even at their resolution of $\sim$22,000, many heavy element features were not unambiguously detected. \citet{sterlingdinerstein08} performed the first large  scale study of n-capture elements in PNe, with a K band survey of [Kr~{\sc iii}] and [Se~{\sc iv}] emission lines in 120 PNe. However, their abundance determinations were uncertain by factors of 2--3, since only one ion of each element was detected (leading to large and uncertain corrections for unobserved ions). Deep, high-resolution optical spectroscopy of PNe enables dramatic improvements to the accuracy of {\emph n}-capture element abundance determinations. The optical region is home to a multitude of {\emph n}-capture element transitions including lines from multiple ions of Br, Kr, Rb, and Xe. Detecting multiple ions of these species allows their abundances to be determined to unprecedented accuracy, thereby providing strong constraints to models of AGB nucleosynthesis and chemical evolution.

Therefore, to improve the accuracy of {\emph n}-capture element abundance determinations in PNe, we embarked on an ambitious observational program aimed at detecting multiple ions of several {\emph n}-capture species in a small sample of about 8 PNe. The data gathered will enable us to compute accurate total abundances for the object of the sample, as well as verify the consistency of current ICFs based on photoionization modelling for future, less in-depth studies. In addition, the comparison of different lines of the same ionization species will enable us to assess the quality of the (as yet poorly-tested) atomic data for heavy elements. Both aspects are crucial to hone nebular spectroscopy into an effective tool for studying \emph{s}-process nucleosynthesis.

In this work, we present the results obtained for the PN NGC\,3918 with UVES at VLT (Fig.~1), The main aim is measuring the abundances of several \emph{n}-capture elements and compare them with theoretical predictions. Nucleosynthesis models predict a correlation between \emph{s}-process enrichment and carbon abundance. As it has been pointed out before, this work is part of an ambitious project oriented, among other things, to test this correlation. In \citet{garciarojasetal15} we present the extended analysis of the work presented in this proceeding.

\begin{figure}[t!]
\resizebox{\hsize}{!}{\includegraphics[clip=true]{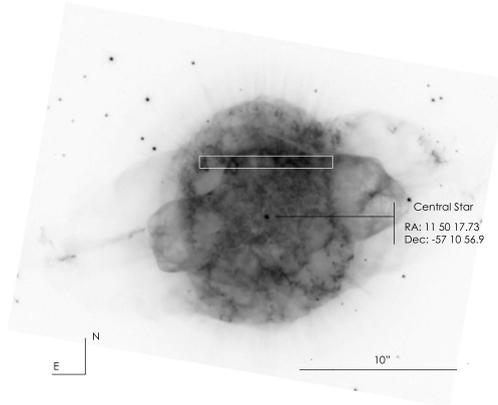}}
\caption{\footnotesize HST H$\alpha$ image of NGC\,3918 showing the UVES slit position.}
\label{slitpos}
\end{figure}

\section{Data and analysis}

We detect, identify and measure more than 750 lines, several of them belonging to different ionic species of \emph{n}-capture elements (Fig.~2). In the cases of very tight blends we used PySSN, a python version of the emission line spectral synthesis code X-SSN \citep{pequignotetal12}, to perform the line fitting (Fig.~3). 
The large number of line intensities allowed us to derive physical conditions using multiple diagnostics. These computations were carried out with PyNeb \citep[v. 1.0.9;][]{luridianaetal15}, a python-based package dedicated to the analysis of emission line spectra. For the calculation of ionic abundances we assumed a three-zone ionization scheme (low-, medium- and high-ionization zone) and each abundance was calculated with the representative physical conditions. Fig.~4 shows several diagnostic ratios for the case of the medium-ionization zone. 
The total abundances of common elements were computed either by adding the ionic abundances or by using ionization correction factors (ICFs) for C, N, O, Ar, Cl, and Ne developed by \citet{delgadoingladaetal14}. 

\begin{figure}[t!]
\resizebox{\hsize}{!}{\includegraphics[clip=true]{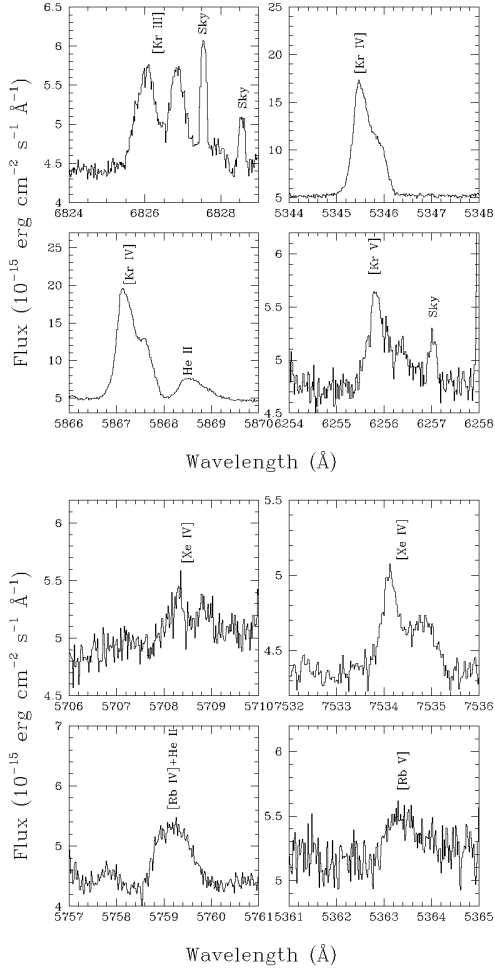}}
\caption{\footnotesize Lines of different ionic species of Kr, Xe and Rb.}
\label{profiles}
\end{figure}

\begin{figure}[t!]
\resizebox{\hsize}{!}{\includegraphics[clip=true]{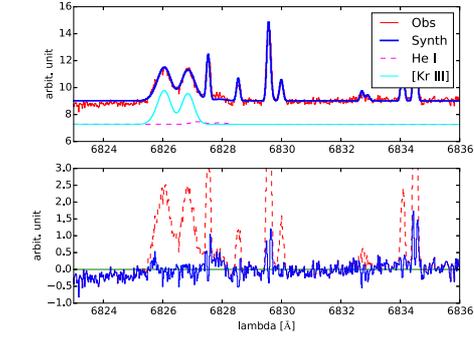}}
\caption{\footnotesize PySSN fitting of the [Kr~{\sc iii}] $\lambda$6826.70 + He~{\sc i} $\lambda$6827.88 feature. Top: observed spectrum (red line) and PySSN fits (cyan: [Kr~{\sc iii}]; magenta: He~{\sc i}; blue: total spectrum; the flux is in arbitrary units). A model of the telluric emission is included in the fit. Bottom: residuals of the fit.}
\label{pyssn}
\end{figure}

\begin{figure}[t!]
\resizebox{\hsize}{!}{\includegraphics[clip=true]{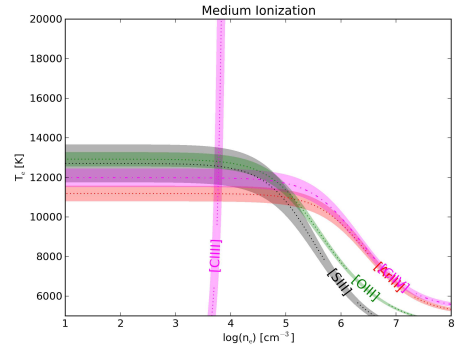}}
\caption{\footnotesize Diagnostic diagram of the medium-ionization zone.}
\label{diagnostic}
\end{figure}

\section{\emph{s}-process enrichments of \emph{n}-capture elements} 

We computed the total abundances of four \emph{n}-capture elements: Se, Kr, Rb and Xe, in order to establish if the central star has undergone a substantial \emph{s}-process nucleosynthesis and convective dredge-up. The enrichment must be measured relative to a reference element which is neither enriched nor depleted in the PN. In the case of NGC\,3918 (a non-type-I PN), oxygen is a suitable element. 
In Table 1 we show the \emph{n}-capture element enrichments represented as [X/O]=log(X/O)$-$log(X/O)$_{\odot}$, where the solar abundances were taken from \citet{asplundetal05b}. For Kr we take advantage of the recent ICFs computed by \citet{sterlingetal15} from detailed photoionization models, and we obtain [Kr/O]=0.60$\pm$0.13, a value well above the threshold assumed by \citet{sterlingdinerstein08}; therefore, we conclude that Kr is strongly enriched in NGC\,3918. For Rb and Xe we cannot draw a definite conclusion, as we do not have reliable ICFs and the adopted values are only lower limits to the real abundances; however, we cannot exclude that these elements are actually enriched. For Se, owing to the difficulties of its detection \citep[see][]{garciarojasetal15}, the value we find is very uncertain. 

\begin{table}
\caption{Total abundance ratios}
\label{abun}
\begin{center}
\begin{tabular}{lc}
\hline \noalign{\smallskip} 
Ratio & [X/O]  \\ [3pt]
\hline \noalign{\smallskip} 
[Se/O] &  0.18$^{+0.17}_{-0.22}$  \\ [3pt]
[Kr/O]  &  0.60$\pm$0.13  \\ [3pt]
[Rb/O] & $>$0.17$^{+0.17}_{-0.22}$  \\ [3pt]
[Xe/O] & $>$0.21$\pm$0.11 \\ [3pt]
\hline
\end{tabular}
\end{center}
\end{table}

\subsection{Correlation with Carbon}

Nucleosynthesis models predict that {\emph n}-capture element enrichments correlate with the C/O ratio, as carbon is brought to the surface of AGB stars along with {\emph s}-processed material during third dredge-up episodes \citep[see e.~g.][]{karakaslattanzio14}. In Fig.~5 we show the correlation we found for a sample of objects with measured kripton enrichments with the C/O ratio. Green triangles represent PNe observed and studied by \citet{garciarojasetal09} and \citet{garciarojasetal12}, and the blue dot is NGC\,3918. The red line is our fit and the blue line is the fit found by \citet{sterlingdinerstein08}, with a larger sample of objects. Our C/O ratios where computed from optical recombination lines, whereas Sterling \& Dinerstein ones where computed from UV and optical collisionally excited lines. In spite of the small size of our sample, both fits are very similar.

\begin{figure}[]
\resizebox{\hsize}{!}{\includegraphics[clip=true]{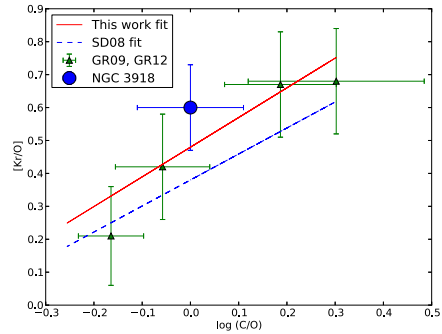}}
\caption{
\footnotesize . [Kr/O] vs. log(C/O) correlation. NGC\,3918 (blue dot) and objects from the sample of \citet{garciarojasetal09} and \citet{garciarojasetal12} with measured [Kr~{\sc iii}] and [Kr~{\sc iv}] lines (green triangles). A least squares fit to the data is shown as a red line. For comparison we also show the fit to the \citet{sterlingdinerstein08} data (dashed blue line).}
\label{C_fit}
\end{figure}
 
\begin{acknowledgements}

We are very grateful to the scientific organisers of the EWASS 2015 Special Session ``AGB stars: a key ingredient in the understanding and interpretation of stellar populations'' for organizing such an encouraging and stimulating meeting. 
\end{acknowledgements}

\bibliographystyle{aa}

\begin{thebibliography}{}

\bibitem[\protect\citeauthoryear{{Asplund}, {Grevesse} \& {Sauval}}{{Asplund}
  et~al.}{2005}]{asplundetal05b}
{Asplund} M.,  {Grevesse} N.,    {Sauval} A.~J.,  2005, in {Barnes} III T.~G.,
  {Bash} F.~N.,  eds, ASP Conf. Ser. 336: Cosmic Abundances as Records of
  Stellar Evolution and Nucleosynthesis p.~25

\bibitem[\protect\citeauthoryear{{Delgado-Inglada}, {Morisset} \&
  {Stasi{\'n}ska}}{{Delgado-Inglada} et~al.}{2014}]{delgadoingladaetal14}
{Delgado-Inglada} G.,  {Morisset} C.,    {Stasi{\'n}ska} G.,  2014, MNRAS, 440, 536

\bibitem[\protect\citeauthoryear{{Garc{\'{\i}}a-Rojas} {Madonna}, {Luridiana}, {Sterling}, {Morisset},
  {Delgado-Inglada} \& {Toribio San Cipriano}}{{Garc{\'{\i}}a-Rojas}
  et~al.}{2015}]{garciarojasetal15}
{Garc{\'{\i}}a-Rojas} J.,  {Madonna} S., {Luridiana} V., {Sterling} N.~C., {Morisset} C.,  {Delgado-Inglada} G.,
   {Toribio San Cipriano} L.,  2015, MNRAS, 452, 2606

\bibitem[\protect\citeauthoryear{{Garc{\'{\i}}a-Rojas}, {Pe{\~n}a}, {Morisset},
  {Mesa-Delgado} \& {Ruiz}}{{Garc{\'{\i}}a-Rojas}
  et~al.}{2012}]{garciarojasetal12}
{Garc{\'{\i}}a-Rojas} J.,  {Pe{\~n}a} M.,  {Morisset} C.,  {Mesa-Delgado} A.,
   {Ruiz} M.~T.,  2012, A\&A, 538, A54

\bibitem[\protect\citeauthoryear{{Garc{\'{\i}}a-Rojas}, {Pe{\~n}a} \&
  {Peimbert}}{{Garc{\'{\i}}a-Rojas} et~al.}{2009}]{garciarojasetal09}
{Garc{\'{\i}}a-Rojas} J.,  {Pe{\~n}a} M.,    {Peimbert} A.,  2009, A\&A, 496,
  139

\bibitem[\protect\citeauthoryear{{Karakas} \&
  {Lattanzio}}{{Karakas} \& {Lattanzio}}{2014}]{karakaslattanzio14}
{Karakas} A.~I., {Lattanzio} J.~C.,  2014, PASA, 31, 62

\bibitem[\protect\citeauthoryear{{Luridiana}, {Morisset} \& {Shaw}}{{Luridiana}
  et~al.}{2015}]{luridianaetal15}
{Luridiana} V.,  {Morisset} C.,    {Shaw} R.~A.,  2015, A\&A, 573, A42

\bibitem[\protect\citeauthoryear{{P{\'e}quignot} \& {Baluteau}}{{P{\'e}quignot}
  \& {Baluteau}}{1994}]{pequignotbaluteau94}
{P{\'e}quignot} D.,  {Baluteau} J.-P.,  1994, A\&A, 283, 593

\bibitem[\protect\citeauthoryear{{P{\'e}quignot}, {Morisset} \&
  {Casassus}}{{P{\'e}quignot} et~al.}{2012}]{pequignotetal12}
{P{\'e}quignot} D.,  {Morisset} C.,    {Casassus} S.,  2012, in {Manchado} A.,
  {Stanghellini} L.,   {Sch{\"o}nberner} D.,  eds, Planetary Nebulae (IAU S283)
  Cambridge University Press Vol.~283, {Planetary Nebulae (IAU S283)}.
pp 470--471

\bibitem[\protect\citeauthoryear{{Sharpee}, {Zhang}, {Williams}, {Pellegrini},
  {Cavagnolo}, {Baldwin}, {Phillips} \& {Liu}}{{Sharpee}
  et~al.}{2007}]{sharpeeetal07}
{Sharpee} B.,  {Zhang} Y.,  {Williams} R.,  {Pellegrini} E.,  {Cavagnolo} K.,
  {Baldwin} J.~A.,  {Phillips} M.,    {Liu} X.-W.,  2007, ApJ, 659, 1265

\bibitem[\protect\citeauthoryear{{Sterling} \& {Dinerstein}}{{Sterling} \&
  {Dinerstein}}{2008}]{sterlingdinerstein08}
{Sterling} N.~C.,  {Dinerstein} H.~L.,  2008, ApJS, 174, 158 

\bibitem[\protect\citeauthoryear{{Sterling}, {Porter} \&
  {Dinerstein}}{{Sterling} et~al.}{2015}]{sterlingetal15}
{Sterling} N.~C.,  {Porter} R.~L.,  {Dinerstein} H.~L.,  2015, ApJS, 218, 25 



\end{thebibliography}

\end{document}